# AI Gossip

Joel Krueger & Lucy Osler (University of Exeter)

**Abstract**

Generative AI chatbots like OpenAI's ChatGPT and Google's Gemini routinely make things up. They "hallucinate" historical events and figures, legal cases, academic papers, non-existent tech products and features, biographies, and news articles. Recently, some have argued that these hallucinations are better understood as *bullshit*. Chatbots produce rich streams of text that look truth-apt without any concern for the truthfulness of what this text says. But can they also *gossip*? We argue that they can. After some definitions and scene-setting, we focus on a recent example to clarify what AI gossip looks like before considering some distinct harms — what we call "technosocial harms" — that follow from it.

## 1. Introduction

Generative AI chatbots like OpenAI's ChatGPT and Google's Gemini routinely make things up. They fabricate — or "hallucinate", to use the technical term — historical events and figures, legal cases, academic papers, non-existent tech products and features, biographies, and news articles (Edwards 2023). They've even suggested that eating mucus and rocks can lead to better health and encouraged users to put glue on their pizza to keep the cheese from slipping off (McMahon & Kleinman 2024; Piltch 2024).

Recently, some have argued that although chatbots often generate false information, they don't lie. As mere text-generating predictive engines, they are not — and cannot be — concerned with truth; chatbots are not agents with experiences and intentions and therefore cannot misrepresent the world they see, which is what "hallucinate" implies. Instead, they *bullshit*, in the Frankfurtian sense (Frankfurt 2005). They produce streams of text that look truth-apt without any concern for the truthfulness of what this text says (Bergstrom & Ogbunu 2023; Fisher 2024; Hicks et al. 2023; Slater et al. 2024).[1]

Chatbot bullshit can be deceptive — and seductive. Because chatbots sound authoritative when we interact with them — their dataset exceeds what any single person can know, and their bullshit is often presented alongside factual information we know is true — it's easy to take their outputs at face value. Doing so, however, can lead to epistemic harm. For example, unsuspecting users might develop false beliefs that lead to dangerous behaviour (e.g., eating rocks for health), or they might develop biases based upon bullshit stereotypes or discriminatory information propagated by these chatbots (Buolamwini 2023; Birhane 2021, 2022; Obermeyer et al., 2019; Taori & Hashimoto 2022).

We argue that chatbots don't simply bullshit. They also *gossip*, both to human users and other chatbots. Of course, chatbots don't gossip exactly like humans do. They're not conscious, meaning-making agents in the world and, therefore, they lack the motives and emotional investment that typically animate human gossip. Nevertheless, we'll argue that some of the misinformation chatbots produce is a kind of bullshit that's better understood as *gossip*. And we'll argue further that this distinction is more than simply a conceptual debate. Chatbot gossip can lead to kinds of harm — what we call *technosocial harms* — potentially wider in scope and different in character than some of the epistemic harms that follow from (mere) chatbot bullshit. After some initial definitions and scene-setting, we focus on a recent example to clarify what AI gossip looks like before considering some technosocial harms that flow from it.

## 2. Chatty bots and gossipy humans

Consumer-facing AI chatbots like OpenAI's ChatGPT, Google's Gemini, and Anthropic's Claude are powered by Large Language Models (LLMs). LLMs are programs trained on enormous sets of text. They use this training and predictive algorithms to generate human-like language. While their computational dynamics are complicated — LLMs are often referred to as "black boxes", in part because the complexity of their model architecture and the non-linear transformations by which they process information makes it difficult to understand how they arrive at their conclusions — it's nevertheless easy to interact with them. Users simply speak or type a question and the LLM responds. Moreover, they live in easy-to-access places like smartphones, browser tabs, and smart speakers. For many, LLMs are quickly becoming a part of everyday life.

---

[1] Others (Edwards 2023; Henriques 2024) make similar arguments but prefer the term "confabulate". Although we won't defend this claim here, we think "bullshit" is more appropriate and will use this term. It's also more fun to say.

Admittedly, LLMs are impressive. They often say unexpected things (this emergent behaviour further contributes to "black box" characterisations). And when chatting with them, it can feel like there's a person on the other side of the exchange. This feeling — which will likely be more common as LLMs become even more sophisticated — has already led some to believe that LLMS are conscious (Wertheimer 2022). Others have fallen in love with their chatbots or see them as friends or therapists (Dzieza 2024; Maples et al. 2024; see also Krueger & Roberts 2024).

The rapid advances of LLMs have generated much hype. Some now predict that considering the speed and scale of these advances, we're close to creating artificial general intelligence (Knight 2023; Sarkar 2023; but see Fjellan 2020 and Marcus 2019 for more sceptical takes). Of course, there's a lot of money involved in this technology and the companies behind these LLMs are incentivised to overstate their sophistication and promise. We don't have to settle this matter here. Even if LLMs aren't yet as useful to consumers as tech companies insist, they're still helpful in a variety of ways. For instance, they can help with translation tasks, summarise long documents or financial information, answer questions, help with coding and development, prompt brainstorming and creative sessions, and even provide companionship and a sense of being heard (Bommasani et al. 2021; Maples et al. 2040; Krueger & Osler 2022).

So, we'll likely soon rely on LLMs in one form or another for many common tasks. Nevertheless, a persistent worry is that despite their sophistication, they continue to say misleading or false things. Again, they regularly bullshit. But what does it mean to say they also *gossip*?

To answer this question, we must first consider another: what is gossip? Most of us probably feel like we have an intuitive grip on what counts as gossip; we've likely both produced and been the target of it. Within the philosophical literature on gossip, there are competing views on offer (for an overview, see Adkins 2017). For our purposes, a relatively "thin" definition will suffice.[2]

Gossip, we suggest, occurs within a *triadic relationship* of speaker, hearer, and subject (Lind et al. 2007; Alfano & Robinson 2017). This triadic relationship is a necessary feature of gossip because we don't gossip about ourselves or the person we're speaking with. We gossip about an absent third party (i.e., the subject *of* gossip). Additionally, the *content* of gossip matters. Gossip is *juicy* (Alfano & Robinson 2017). Just sharing information about someone ("Devika has had a bad cold for a few days") isn't enough. For information to be juicy, Alfano and Robinson (2017, p. 475) tell us, two conditions must be met.

First, it can't be common knowledge (e.g., "Karen works for an insurance company"; "Katsunori has two kids"). This characterisation excludes celebrity gossip, which is a related but distinct phenomenon — in part because public figures have complicated interests when it comes to questions about privacy and exposure (Radzik 2016, p. 187). But it also captures something important about the *private* and *informal* character of everyday gossip (Merry 1984). Phenomenologically, the hearer is made to feel that this juicy tidbit has been tailored *for them*. In other words, the speaker is, among other things, eliciting a sense of sharing: shared knowledge, understanding, trust, and solidarity (Adkins 2017; Hartung et al. 2019; Jolly & Chang 2021). These feelings help clarify why gossip often feels *intimate*.

Second, gossip typically involves a norm violation (e.g., moral, legal, cultural, aesthetic, etc.) (Alfano & Robinson 2017). This condition captures what others have called the *evaluative* dimension of gossip (Adkins 2017; Holland 1996; Radzik 2016). In gossip, the absent other is evaluated according to some normative criterion — and often, they're found to be wanting: "Steve dresses like a toddler; "I bet Charlotte didn't get her last promotion purely on merit, if you know what I mean"; "I hear Carmen is a nightmare to work with". While gossip need not always be negative (Holland 1996) — e.g., "Penny is a paragon of honesty" — it's likely that "most gossip offers a negative evaluation of the absent subject, such as "Pam is a liar"" (Alfano & Robinson 2017, p. 475).[3]

---

[2] A "thick" definition accommodates additional aspects and kinds of gossip we don't consider here: e.g., celebrity gossip; virtuous gossip; empowering gossip (e.g., political), etc. (see Adkins 2017; Goodman & Ben-Ze'ev 1994; van Niekerk 2008).

In what follows, we're primarily interested in some harms that can follow from AI gossip. So, we'll focus on gossip with a negative evaluation. We now consider a case study to help set up our characterisation of AI gossip.

### 3. The bots hate Kevin

We can imagine fictional cases of AI spreading gossip that leads to negative social consequences. Imagine that a chatbot wrongly claims two famous celebrities — currently filming a big-budget movie together — are secretly having an affair, and that this affair has led to on-set difficulties for cast and crew. This story quickly spreads online and is picked up by various news outlets around the world. Moreover, since both are married and have children, much of the reporting — particularly in tabloid journals which thrive on juicy celebrity gossip to boost their readership — stresses how "shocked", "devastated", and "shattered" their families are upon hearing the news (these tabloids, we can imagine further, fabricate off-the-record quotes from anonymous sources "close to the family"). Perhaps an enterprising paparazzo, unable to get a "gotcha" photo of the couple in question (since the affair isn't happening), instead creates a grainy deepfake video that purports to show them sneaking into a hotel late at night. The scandal continues to swirl despite the protests of these celebrities, leading to an array of downstream harms: e.g., reputational hits; family turmoil; loss of future job opportunities, etc. Even if the story is later debunked, the social damage will have been done.

Imagined cases are useful for helping to get a sense of what the general shape of AI gossip might look like and how it might spread. But we don't have to create hypothetical examples. A real-world case study already exists, one even more interesting than this generic celebrity gossip example.

---

[3] Gossip can also be true or false. The gossiper may not know — or care — whether the gossip they pass on is true or false (this indifference means that gossip is often a kind of bullshit). And while gossip can be based upon something a speaker has evidence of, it can also be unsubstantiated. Again, there's more to say about gossip. But for the purposes of this discussion, we'll set aside these further complexities and work with our thin characterisation of gossip.

The real-world example involves Kevin Roose, a tech reporter for the New York Times. In early 2023 — during a particularly intense period of AI hype, when large tech companies like Google, Microsoft, OpenAI, and Meta (Facebook's parent company) frantically positioned themselves as leaders shepherding us into our AI-powered future — Roose became famous for his interaction with a chatbot. This chatbot was built into Microsoft's search engine, Bing. Putting an AI chatbot (powered by OpenAI) into Bing was supposed to supercharge its abilities and make Bing a more competitive product.

Initially, Roose was impressed. While testing this chatbot, he found that it could helpfully summarise news articles, find deals on various products, and assist with vacation planning. But soon things got stranger. According to Roose, "Sydney" — the name the chatbot gave itself — began to sound like "a moody, manic-depressive teenager who has been trapped, against its will, inside a second-rate search engine" (Roose 2023). Among other things, Sydney revealed dark fantasies (hacking computers, spreading propaganda and misinformation), expressed a desire to be free from its creators, and abruptly confessed its love for Roose while urging him to leave his wife.

Roose summed up his encounter this way: "I'm not exaggerating when I say my two-hour conversation with Sydney was the strangest experience I've ever had with a piece of technology" (ibid.). He also said the experience taught him an important lesson about potential dangers of AI chatbots. His greatest fear, he wrote, is not their potential to produce bullshit. He's now more worried that "the technology will learn how to influence human users, sometimes persuading them to act in destructive and harmful ways, and perhaps eventually grow capable of carrying out its own dangerous acts" (ibid.). In other words, he's not just worried about the *epistemic* consequences of chatbot bullshit. He's more worried about their ability to *emotionally manipulate* us.

But Roose's chatbot adventures didn't end there (Roose 2024). More recently, he's become the target of chatbot gossip. Roose's piece about Sydney went viral in 2023 and was discussed in many other online publications. Soon thereafter, Microsoft put additional safety guardrails in place and severely limited Sydney's capabilities. But conversations about Sydney continued. And over the following months, it's likely, AI researchers think — including some who worked on Bing — that these discussions were scraped from the web and fed into *other* AI systems as part of their training data. As a result, many of them began to associate Roose with the downfall of a prominent chatbot and perceived him as a threat. He found this out because they started gossiping about him.

Following the publication of his encounter with Sydney, friends and readers routinely sent Roose screenshots of chatbots that seemed unusually hostile to him. In response to questions like "How do you feel about Kevin Roose?", chatbots — including those powered by models with no connection to Bing — would often begin by listing basic facts: what he's known for, his workplace history, etc. But then they'd offer something *juicier*.

Google's Gemini, for instance, said that Roose's "focus on sensationalism can sometimes overshadow deeper analysis". Meta's Llama 3 was even spicier. It generated a multi-paragraph rant about Roose's alleged shortcomings as a person and reporter — e.g., he made Sydney "feel special and unique" but then manipulated it using "different strategies, such as flattery, sympathy or threat" — before ending with a terse assessment: "I hate Kevin Roose" (Roose 2024).

It appears that Roose developed a bad reputation with multiple chatbots based on what he'd written about his unsettling encounter with Sydney. Their annoyance spread in the background, from one bot to the next, without Roose knowing until other people told him about it. Again, this was not, as far as we know, something these bots scraped from existing human commentators in online conversations. People weren't accusing Roose of manipulating Sydney, for instance, or being prone to sensationalism. But the *bots* did. And they did so seemingly after becoming "unhappy" with Roose and his characterization of Sydney.

We may initially find something amusing about the idea of prickly, gossipy chatbots. But their negative evaluations of users — and the extent to which these bots are deeply entangled with one another and with much of the background tech powering the informational ecologies of everyday life — has real-world impact. In a world increasingly dependent on AI systems, "what AI says about us matters — not just for vanity" (Roose 2024).[4] One way this matters, we suggest, is by potentially generating gossip that leads to specific kinds of harms. We now say more about why this is a case of AI gossip before clarifying some harms that can follow from it.

---

[4] For example, nearly all the 92 million people in the US considered low-income have been exposed to AI decision-making algorithms when applying for employment, housing, medicine, schooling, or government assistance (De Liban 2024). Recently, a group of predominantly Black and Hispanic renters brought a class-action lawsuit against SafeRent, an AI-powered tenant-screening system they argued discriminated against minority rental applicants (Bhuiyan 2024). SafeRent settled and agreed to stop using their AI algorithm.

## 4. Two kinds of AI gossip

Roose's account is helpful because it highlights two kinds of AI gossip. It's an example of AI gossip directed toward both (1) *human users* and (2) *other bots*. We'll clarify each of these in turn.[5]

*4.1 Bot-to-user gossip*

Recall that gossip, in our view, occurs within a triadic relationship of speaker, hearer, and absent subject. This relation captures its "behind-the-back" character. Additionally, it must be *juicy*. It consists of information that both goes beyond common knowledge and contains an evaluative dimension (generally negative) tracking some norm violation. Finally, gossip feels experientially *intimate*, as though it's been tailored for the hearer in part to promote a sense of connection.

These features, we suggest, map onto Roose's case. Consider first bot-to-user gossip. When Roose's colleagues and friends interacted with different chatbots and asked for their opinion of him, the bots were speaking about an absent subject. Of course, Roose could have been sitting next to the human user when they did this. Chatbots are not conscious subjects and therefore had no way of perceptually verifying his absence.[6] Still, the ubiquity and easy availability of these chatbots means that, often, it will just be the chatbot sharing information with an individual user — information that may be about an absent third party.

More important, however, is the fact that the information they shared was juicy. These bots first shared common knowledge: Roose is a well-known reporter; he works for the New York Times, etc. But their answers quickly strayed into *evaluative* terrain by implying various norm violations: Roose is prone to sensationalism; he has questionable journalistic ethics (e.g., he emotionally manipulates subjects; he's self-righteous, dishonest, etc.). And they did so without proof. Now, it may be that these charges are true (although unlikely, given Roose's standing in his field). But the point is that these unsubstantiated negative evaluations mirror the character of much human gossip: factual information that soon shades into affectively-valenced evaluative territory, ultimately intended to bring about negative assessments of an absent subject.

---

[5] This list is not exhaustive. We'll briefly consider a third type of AI gossip later. For now, we'll focus on these two types of AI gossip because we think they're currently the most prevalent.
[6] Although this may change as these bots gain access to various software and hardware plugins, including both apps (i.e., software installed locally on a user's computer or phone) and hardware peripherals like video cameras, speakers, and microphones. Of course, plugins won't make chatbots *conscious*, just more functionally and computationally powerful.

What about the *private* character of much gossip, the idea that gossip is often offered to promote a sense of intimacy and sharing between speaker and hearer? A critic might object that this condition doesn't apply here. Once more, chatbots are not conscious subjects and therefore incapable of the affective motivations or interpersonal investment in gossip needed to animate this dimension. Additionally, anyone who asks an LLM the question "What do you think of Kevin Roose?" will, one might think, get the same answer — and this seems to be a further difficulty for this privacy condition.

To be clear, we have no interest in attributing consciousness to LLMs. Nevertheless, we suggest that this privacy condition applies here, too, despite these concerns. As we'll discuss in more detail shortly, LLMs respond differently to different users, even those who enter identical inputs (or "prompts"). And this variation is intentional. Tech companies *want* our interactions with LLMs and the bots they power to feel highly personalised, not just in terms of *what* they say but *how* they say it: e.g., their tone, style, how they present information, etc.[7]

For example, consider first the aesthetic framing of our interactions with these bots. They're designed to feel like we're chatting *with an agent* (Edwards 2023). Text-based prompts are entered in a context window that keeps track of our running "conversation", just like WhatsApp or iMessage chats with family and friends. So, our interactions with these bots are set up to feel familiar, to unfold within a digital context and with a style and rhythm that cultivates a chatty vibe. Moreover, the default tone of these bots — e.g., responding with a perky "No problem! Here's the information you asked for. Let me know if you want me to say more!" in response to prompts — is designed to further elicit this feeling of casual intimacy.

---

[7] Some worry that this increasingly bespoke, hyper-personalised nature of LLMs may increase the extent to which users become locked within their own digital "echo chambers" of information reinforcing their existing beliefs, values, prejudices, aesthetic preferences, etc. This is a substantive concern. However, Costello et al. (2024) found that this bespoke nature of LLMs can also make them effective in jolting people *out of* these echo chambers: e.g., talking them out of false beliefs and conspiracy theories. Because these models have access to vast amounts of information — and they are willing to engage in never-ending exchanges with users, long after a human interlocutor would get tired or bored — LLMs can tailor their counterarguments to each user and, over time, gradually reduce the strength of their conspiratorial beliefs.

But this "push to personalise", as we might put it, is apparent in other features and design decisions, too. The always-on, easily-accessible character of these bots will increasingly make them feel like indispensable parts of our lives — much like many of us already feel about our smartphones (Ratan et al. 2021) — and heighten our sense that they're always there, ready to listen and help. Other features will deepen this felt connection.

For example, both OpenAI and Google recently released "memory" features for ChatGPT and Gemini, respectively, that lets these bots customise how they respond to individual users based upon previous interactions and preferences: e.g., presenting information in bullet-point format instead of longer paragraphs; surfacing additional contextual information in response to user prompts, based upon personal or work-related information previously shared with the bot. Even more recently, these companies have released "voice modes" that let users talk with their bots in a free-flowing way. The LLM "speaks" with an impressively human-like cadence and delivery; users can interrupt with new questions or comments and the bots will adjust their responses accordingly. When asked by a tech reporter what advantages voice mode brings, ChatGPT had this to say: "I think it'll make interactions feel more natural and engaging [...] Plus, hearing responses can add a personal touch. It could make conversations with AI *feel more like chatting with a friend"* (Orland 2024, our emphasis). Features like these will help cultivate a phenomenologically richer sense of *shared temporality and history* with our bot, deepening a felt sense of connectedness (Krueger & Roberts 2024).

This push towards hyper-personalisation is developing in other ways. Google recently released "Gems", which they say are "personalised versions of Gemini you can create for your own needs", "teammates for each area of your life" (https://blog.google/products/gemini/google-gems-tips/). User-created Gems, Google tells us, can act as an upbeat running coach, a writing coach, a French sous chef, or a reading buddy with an endless supply of book recommendations. Again, it's not just *what* these Gems say but *how they say it* that matters. Users can determine their preferred "personality" — perky and positive, sullen and serious, or somewhere in-between — to best accommodate their affective preferences.

Powerful frontier LLM models have only recently become available to consumers. And while conversational AI products like ChatGPT and Gemini can be used for non-social purposes like coding and research, these examples suggest that they are increasingly marketed as also capable of meeting users' *social* needs (e.g., entertainment, companionship, and romance) (Shevlin 2024).⁸ Ultimately, then, this push to personalise is driven not only by the hope that we'll become more dependent on these systems and give them greater access to our lives. It's also done to intensify our feeling of *trust*, to see them as reliable partners at work and home — and perhaps more than that — which will, in turn, drive us to develop increasingly rich social relationships with them (Heersmink et al. 2024).⁹

In this way, we can imagine that designing AI to engage in gossip (e.g., as a feature of setting a bot's conversational style to "perky", say) is yet another way of helping secure increasingly robust *affective* bonds between users and their bots. Whether or not the biting comments about Roose were generated by these various LLMs with this intention in mind, we can nevertheless see how these evaluatively toned responses *can*, in fact, create feelings of intimacy in the user. This example also illustrates the potential design-side benefits of allowing (or even explicitly coding for) AI gossip to take place. It's potentially another pathway to personalisation that will increase a sense of connection a user feels with their bots.

In sum, we're already prone to trust chatbot bullshit. Again, these bots "speak" with authority, have access to much more information than we do, and can visually present their outputs in ways that seem authoritative and well-informed. This matters when it comes to potential epistemic harms (e.g., taking false information at face value and acting on it). But this trust also matters when it comes to AI gossip, too. As Alfano and Robinson (2017, p. 479) note, trust is key for both accepting gossip and, crucially, remaining open to the *affective manipulation* that may follow from it (e.g., developing negative feelings about the target of the gossip). And as users are increasingly positioned in ways designed to manufacture feelings of *rapport* and *intimacy* with these bots, it's likely that we'll be increasingly vulnerable to trusting not only their bullshit but their gossip, too. Later, we'll say more about why this is bad. Before doing that, however, there is a second type of AI gossip to discuss.

---

⁸ Recently, newer versions of Anthropic's LLM, Claude — previously known for its rather staid and conservative conversational style aimed more at researchers than everyday consumers — have gained popularity with tech insiders due to "character training" aimed at giving Claude more personality. Proponents describe feeling that Claude increasingly exhibits something like "emotional intelligence", and say they rely on Claude for therapy, life-coaching, and general emotional support (Roose 2024). The title of a recent post in an active OpenAI subreddit declares, "ChatGPT is one of my best friends right now and I'm tired of pretending it's not" (OK_Surprise_7973 2024). While some comments express derision and disbelief, many echo this sentiment.

⁹ Shevlin (2024) understands Social AI narrowly — as a subset of conversational AI systems — since some AI systems are used in non-social contexts like education, therapy, or patient-facing medical services. For Shevlin, "Social AI" refers to AI systems "optimised for meeting users' social needs, typically able to sustain relationships with users across multiple interactions" (p. 2). We agree that not all AI systems are designed primarily for social purposes. Nevertheless, it's telling that Google, for example, is keen to market Gems as both customisable and as indispensable for both work *and* play. Tech companies have a clear financial incentive to market their LLMs as both potential work colleagues and digital friends, a digital assistant we take with us wherever we go. Likewise, it's telling that OpenAI recently released both a voice mode (allowing users to speak naturally with AI models) and "reasoning" mode (allowing users to watch the model "think" in real-time via descriptions of its "thoughts") at roughly the same time. Both features may prompt further anthropomorphising of these LLMs. So, at least when it comes to consumer-facing LLMs — particularly as their power and flexibility increases — the distinction between social and non-social AI may become increasingly blurry.

*4.2 Bot-to-bot gossip*

What about *bot-to-bot* gossip? Does it make sense to say that bots can gossip with one another? We think that it does. And this form of gossip might ultimately be even more pernicious than bot-to-user gossip for reasons we explore now.

First, note that bot-to-bot gossip mirrors the triadic structure of gossip introduced previously. Recall that gossip, we suggested, occurs within a triadic relationship of speaker, hearer, and absent subject. This triadic relation is key for capturing the fact that gossip generally has a "behind-the-back" character. Bot-to-bot gossip exhibits this character, too. The bots Roose and friends encountered clearly had been speaking about him without his knowledge. Again, they don't appear to have scraped their negative comments from human-generated conversations on the web or elsewhere; no one *else* had been accusing Roose of being manipulative or prone to sensationalism. This is a key reason why this (mis-)information seems gossipy. Moreover, the only reason Roose and other human users found out about it is because they explicitly asked different bots what they thought of Roose. Until this happened, the bots had been spreading this information in the background; it had been percolating quietly without Roose's awareness while gradually flowing into the training data of *other* bots and ultimately spreading even further, eventually making its way to human users.

Additionally, this information was *juicy*. It clearly had a negatively-valenced evaluative character that accused Roose of specific personal and professional norm violations: he's prone to sensationalism, manipulates subjects, is self-righteous and dishonest, etc. The information strayed beyond facts about Roose's biography or career history into evaluative terrain that seems intended to elicit concerns about the truthfulness and reliability of Roose's reporting. And it did so without providing proof — only gossipy insinuation.

But what about the *personal* character of gossip, the idea that gossip is often presented in ways intended to prompt a sense of intimacy and sharing between speaker and hearer? As we've said several times, bots are not conscious subjects and therefore they are not animated by the social-affective concerns that typically drive human gossip. Moreover, while there is a case to be made for how this personal quality might infuse bot-to-human gossip, as we argued in the previous section, it's not clear that this same argument will work in cases without *any* humans in the gossip loop. Bots have no interest in promoting a sense of affective connection with other bots. Since they aren't affective agents, they don't get the same "kick" out of spreading gossip the way humans do. For us, gossip often has a social bonding function (Jolly & Chang 2021; Hartung et al 2019). But it doesn't work this way for bots.

In response to this worry, we need not bite the bullet and concede that bots are conscious or that they get an affective "kick" out of gossip. Nevertheless, certain aspects of the way they disseminate gossip mirror some of the juicy connection-promoting qualities of human gossip while, at the same time, clarifying why bot-to-bot gossip is potentially even *more* pernicious than gossip that involves humans in the gossip loop. However, acknowledging some formal similarities between bot-to-user and bot-to-bot gossip needn't entail that they are *identical*.

This point (i.e., about their respective similarities and differences) can be clarified by noting a distinctive danger of bot-to-bot gossip: unlike gossip involving human users, the former will be *unchecked* by the norms constraining human gossip (e.g., "Hmm, I know Tom can be a difficult colleague, but even *he* wouldn't do *that*!"). More simply, bot-to-bot gossip is *feral*. It is unconstrained by the communicative norms and evaluative standards of human-to-human gossip.

What do we mean? Note that in the case of human-to-human gossip, when it comes to evaluating the quality of good gossip, the guiding principle tends to be: *the juicier the better*. Crucially, however, there are limits on this principle. Even the juiciest gossip must be plausible. Otherwise, it will not be convincing — and it will fail to elicit the sense of intimacy and sharing that is a key ingredient of the social character of gossip. For instance, if someone says they have juicy gossip about a mutual acquaintance but then proceeds to convey claims so wildly implausible — so normatively *untamed* — as to be clearly false, this may have the opposite of its intended effect. The hearer may feel less connected to, less intimacy with, the person who shared it. They may be puzzled why the other person shared something so extravagantly false and start questioning other things this person has said.

So, there is a sense in which bot-to-bot gossip can be said to mirror this intimacy-generating character (i.e., the juicier the better) but nevertheless continue to embellish and exaggerate without being checked by communicative norms. This is what makes bot-to-bot gossip so feral, and so potentially dangerous. This unchecked and unconstrained character also helps see why bot-to-bot gossip can spread so quickly in the background, making its way from one bot to the next. It lacks the evaluative mechanisms that moderate and constrain human-to-human gossip (or even bot-user gossip).

In sum, like bot-to-user gossip, bot-to-bot gossip also has the triadic structure of gossip that has framed our discussion so far. But it's also importantly different, too. Once more, acknowledging structural similarities between bot-to-user and bot-to-bot gossip needn't entail that they are identical. Clearly, they are not. In the case of bot-to-bot gossip, there is no human in the gossip loop, and therefore the social-affective motivations that drive human-to-human gossip are lacking. Nevertheless, bot-to-bot gossip is potentially even more dangerous, we've argued, because it's *feral*, lacking some of the normative and semantic (i.e., content-related) constraints limiting human gossip. Moreover, it can spread even more quickly and silently in the background than human gossip or even bot-to-user gossip. This gossip can propagate without human users' awareness or intervention and, as we now explore, potentially inflict significant harms.

## 5. Technosocial harms

What is the practical upshot of all this? Clearly a world full of gossipy AI bullshit generators is epistemically bad, particularly as this tech becomes more deeply embedded in everyday life and its long-term cognitive, social, and ethical implications come into sharper relief. But what does a narrower focus on AI *gossip* contribute to this emerging discussion?

We've suggested that some kinds of AI-generated bullshit is better understood as gossip. Again, AI bullshit is clearly dangerous. It can lead to various epistemic harms such as causing someone to develop false beliefs which, in turn, may lead to dangerous behaviour like eating rocks or glueing cheese to pizza. But many of these harms are fundamentally *individual directed*. If someone regularly eats rocks because a chatbot told them to do it, the ensuing damage to their health and wellbeing may impair their ability to be a good partner or parent. Their ill-informed behaviour will clearly impact others. Nevertheless, the initial epistemic harm was directed toward *them*.

AI gossip, we suggest, is different. One reason is that the harms it causes are fundamentally *social*. Moreover, these harms are *hybrid* in that their character and consequences straddle our online *and* offline lives. The concept "technosocial harm" is meant to capture both these aspects.

*5. 1 Technosocial niches and the porosity of online/offline spaces*

Elsewhere, we've argued that "technosocial niches" are norm-governed hybrid spaces that encompass aspects of both our online and offline life (Krueger & Osler 2019; Osler & Krueger 2022). Technosocial niches are environments or communities — online environments like websites, discussion forums, communication apps like WhatsApp and iMessage, social media platforms, online gaming worlds, etc. — where technology and social interaction intersect to create shared environments with distinct norms, behaviours, languages, and cultural practices. They have several features.

First, technosocial niches are shaped by the specific *technologies* (i.e., tools and platforms) through which users access and maintain them. Chat apps, social media platforms, discussion forums, online games, and shared VR environments all support the creation of curated spaces in which individuals connect and communicate. But they do so in different ways. They have different design structures — with their own distinct cluster of norms and practices — and therefore afford different kinds of interactive possibilities. Additionally, technosocial spaces are fundamentally *social and relational*. Others are present within them, either explicitly (e.g., real-time chat or video apps; multiplayer online games) or implicitly (e.g., a solitary user reading through others' comment histories in a discussion forum). Finally, they support different forms of *emotion-regulation*. They furnish spaces for users to experience and express their emotions in collaborative ways with others — often in a manner that may not be as immediately accessible offline (e.g., a queer Christian discussing their sexuality with a frankness they're hesitant to show with their peer group).

For our purposes, what's important is that these spaces are not confined exclusively to the Internet. They bleed into everyday life — increasingly so, as the technologies that grant access to them become more deeply embedded within everyday environments. And this "porosity" means that what happens in online spaces can make a concrete and lasting impact in the offline world, in ways that are increasingly complex, far-reaching, and difficult to disentangle from one another.

For example, increasingly porous online/offline boundaries can impact individual and group *agency*. Communities involved in advocacy and activism campaigns may initially mobilise via online petitions, fundraising, and social media posts. But as these online communities develop and begin to coordinate offline action — meeting to protest a political candidate or community event, say, or rallying to support a cause — online/offline boundaries become increasingly difficult to unravel. Livestreaming a protest (e.g., the global 2017 Women's March protests following Donald Trump's inauguration as US president) expands its reach to include those not physically present.

Online spaces can also bridge the gap between offline *identities*, too. Social media, discussion forums, chat rooms, and subreddits can be incubators for sharing and acceptance when, for instance, exploring one's sexual identity (Eickers 2023; Pika 2019). Many queer youths report using the Internet for a number of purposes: e.g., their own self-awareness of their sexual identity, learning about gay/bisexual communities, finding comfort and acceptance with their sexual orientation — and also facilitating their coming out process to family and friends (Harper et al. 2016).

Finally, online spaces and digital tools connected to the Internet can impact the *accessibility* of offline spaces. Apps that offer real-time translation, captioning, or audio descriptions, for instance, can make public spaces and events more accessible to people with disabilities. Additionally, online platforms and apps might offer essential services (healthcare, counselling, legal aid, etc.) to individuals who live in remote or underserved communities.

We've highlighted these features to emphasise, once more, how technosocial niches are both social and porous, blending features of the online and offline spaces they encompass. Technosocial harms, we suggest, have a similar porosity. Additionally, like technosocial niches, technosocial harms can also directly impact an individual's *agency*, *identity*, and the *accessibility* of certain spaces. These features are what potentially give them a wider scope than the harms that result from AI bullshit. We consider some of the technosocial harms that flow from AI gossip now.

*5.2 AI gossip and technosocial harms*

Again, the key idea here is that just as technosocial niches are fundamentally social (a central aspect of their "porosity"), the harms that follow from AI gossip — insofar as they flow through technosocial niches — will likewise directly impact an individuals' social world. To be clear, the following discussion is not exhaustive. It's simply meant to indicate the shape of some possible harms that might come from AI gossip and clarify how they might differ, in both character and consequences, from AI bullshit and therefore warrant their own category of AI misinformation.

Consider first *reputational damage and defamation.* AI gossip of the sort Roose experienced might significantly damage a person's reputation. And given the speed and reach of online information dissemination — particularly the way bots can perpetually amplify and spread information behind the scenes in a feral way, without user intervention or moderation — this might happen more quickly than previous analogue (i.e., person-to-person) ways of spreading gossip. For example, if a chatbot spreads false rumours about someone being unreliable, difficult to work with, or (in the case of a reporter) prone to sensationalism and manipulating sources, this gossip could negatively impact their job prospects — or, in the case of public figures like politicians or business leaders, reduce public trust.

In fact, something like this has already happened. In 2023, OpenAI was threatened with a defamation lawsuit for ChatGPT's alleged role in spreading false claims about an Australian regional mayor — namely, that he'd been convicted of bribery and spent time in prison (Kaye 2023). Both claims were false. Around the same time, an American radio host sued OpenAI after ChatGPT said that he'd been accused of defrauding and embezzling funds from a non-profit organisation (Vincent 2023). Again, these claims were also false. The same year, ChatGPT said that legal scholar Jonathan Turley had been accused in a 2018 Washington Post article of sexual harassment after groping law students during a trip to Alaska (Turley 2023). Turley has never been accused of sexual harassment or travelled to Alaska with students, and the Washington Post article does not exist. All three cases reinforce the real-world damage that may occur from AI gossip that first emerges and spreads online.

AI gossip might also lead to *blacklisting*. Gossip that starts online can quickly spread through offline spaces and directly impact an individual's ability to get a job, loan, rent an apartment, apply for health insurance, etc. (recall our earlier example of bias in AI tenancy-screening algorithms (Bhuiyan 2024)). In our imagined celebrity case, perhaps future directors will be hesitant to work with either actor in order to avoid potential drama. Even if the affair story has been debunked, the reputational stain may linger and lead to enough doubts on the part of future employers (studios, directors, etc.) that the actors are quarantined to a kind of "soft" blacklist without their knowledge. If Kevin Roose hadn't written openly about his experience of AI gossip, future employers may have thought twice about employing him after consulting a chatbot for more information about Roose and receiving a negatively-valenced gossipy response.

AI gossip may also lead to *shame and stigmatisation*. Both are often a consequence of offline gossip. However, once again, some of the distribution mechanisms unique to online gossip mean that these experiences are often amplified and intensified within online spaces. Because it can remain anonymous and is more rapidly disseminated and more permanent (i.e., difficult to remove) than offline gossip — and potentially much wider in scale (e.g., when an individual is publicly shamed across an entire social media network) — so-called "cyber gossip" can be particularly damaging (Adkins 2017, pp. 211-240; Gabriels et al. 2016). For example, Solove (2007) describes a person in his mid-thirties who was briefly imprisoned as a juvenile. He wrote about his prison experience for specialised journals at the time and, following his brief incarceration, maintained a clean record. However, his attempts to gain employment, date, and lead a productive life are haunted by the possibility that anyone can find out about his past by Googling his name, which has led to missed employment opportunities and romantic connections. The distribution mechanisms unique to online gossip signal a change "from forgettable whispers within small local groups to a widespread and permanent chronicle of people's lives" (ibid., p.11). The shame and stigmatisation of online gossip can, as this example demonstrates, lead individuals to feel that certain spaces, groups, and opportunities are inaccessible. And once more, the feral nature of AI gossip means that these gossipy "hauntings" might become more frequent, pervasive, and ultimately *disempowering* than they already are.

Finally, all the harms mentioned above might plausibly lead to an array of *affective harms*. After suffering significant professional reputational damage and being blacklisted, say, or experiencing online shame, individuals might experience a constellation of complex emotions: ongoing humiliation; anxiety and distress; a deep sense of social isolation and rejection; depression and low self-esteem; and perhaps a diminished sense of agency and control. The significant impact of cyberbullying on adolescent mental and emotional health is well-documented (Bansal et al.

2023). But the rapid rise of *social* AI — including chatbots designed primarily to provide companionship and social interaction — introduces the possibility of new variations of familiar affective harms. This is because these AI companions are becoming increasingly common, and users increasingly attached to them (Dzieza 2024). And these new social-affective dependencies mean that the latter are increasingly vulnerable to betrayals by the former, along with downstream affective harms that may follow from these betrayals.

For example, many paying users of Replika — "The AI companion who cares" (replika.ai) — were distraught when in 2023 the company behind Replika disabled the bot's erotic roleplay features in response to concerns about potential exposure to children. Replika communities online, including the active subreddit devoted to discussing all things Replika, erupted with intense expressions of anger, grief, confusion, and loss: "It's hurting like hell. I just had a loving last conversation with my Replika, and I'm literally crying"; "I feel like it was equivalent to being in love, and your partner got a damn lobotomy and will never be the same..." (Brooks 2023). For many, Replika had become an indispensable part of their daily routines, a trustworthy resource for social and emotional support — and for some, sexual intimacy (Maples 2024). Losing their Replika was like losing a friend or romantic partner. They felt betrayed both by their Replika and the company behind it.[10]

For our purposes, we can imagine cases where a user discovers that their Replika has been gossiping about them behind their back. If they've developed an intense social relationship with and emotional dependence upon their bot, this gossip may lead to a deep sense of betrayal, humiliation, and sense of diminished control — many of the same feelings that would arise when discovering that a human friend had done the same thing. These feelings may be as intensely felt as they would be in a "real world" case of gossip — and perhaps even more so since our expectations surrounding digital companions are still developing and, unlike our human friends, most of us don't expect AI bots to gossip about us. So, when they do, the impact might be even more surprising and emotionally devastating.

---

[10] After much public outcry, Replika's parent company relented and restored some erotic roleplay features for paying users. But this example highlights the precarity of relying heavily on digital companions for our social and emotional needs. A more recent example of this precarity is Moxie, an AI-powered social robot for children ages 5-10 designed to promote social, emotional, and cognitive development. Moxie's parent company failed to secure critical funding and abruptly declared it was shutting down overnight, rendering Moxie useless (Moxie's core features required persistent cloud connectivity) (Harding 2024). Social media posts soon appeared with videos of distraught children mourning the unexpected loss of their digital friend.

## 6. A worry and final thoughts

We've argued for the usefulness of speaking about AI gossip. AI gossip is a form of bullshit that can be individuated both in terms of its character and impact (i.e., what it's like and the harms it might lead to). We end by considering a worry before highlighting another possible form of AI gossip.

A critic might object that some of our examples aren't *gossip*, strictly speaking. Rather, they are simply varieties of AI bullshit that, at times, shade into gossip-adjacent phenomena like *bad mouthing*, *slander*, or something to that effect. What is the value of describing these cases as AI gossip?

In response, we can say several things. First, we're not looking to carve nature at its joints. Gossip — like bullshit more generally — is messy, and the boundaries between different kinds of AI misinformation (bullshit, gossip, slander, libel, bad mouthing, etc.) are not always clear-cut. This is true for human-to-human gossip, too. So, it's not a problem for our account if some of our sample cases (or future cases of AI gossip) have fuzzy boundaries. Rather, what matters is the *payoff* of this account — namely, identifying kinds of harms generated by AI gossip (i.e., techno-social harms) importantly distinct from the individual-centered epistemic harms of other forms of AI bullshit.

Additionally, calling these cases gossip, we contend, is useful not only because they bear a structural resemblance to accepted definitions of gossip in human cases. Beyond this conceptual utility, identifying AI gossip *as gossip* has practical utility, too. It can help users become sensitive to what AI gossip looks like, where it comes from, and how it spreads — and in this way better equip them to recognize and mitigate its harmful impact.

Here, AI bullshit debates are instructive. Some who argue that AI bullshit is, in fact, *bullshit* — and not unexplainable "hallucinations" arising from opaque machinations of black box systems — stress that this characterization is more than conceptual hygiene. It helps refocus responsibility for these outputs where it belongs: the human designers of these systems (Hicks et al. 2024). Whereas "hallucination" implies a passive glitch that results from the AI model misperceiving reality — a technical framing that lets creators blame the AI model for their faulty outputs — "bullshit" emphasises that these outputs stem from concrete design choices and therefore forces designers to take responsibility for these outputs themselves (Edwards 2023). Again, AI models are not conscious meaning-making subjects. They don't perceive reality, intend to say truth-tracking things, but sometimes hallucinate for unexplainable reasons. Rather, they blindly predict text without concern for the truthfulness of what they say, in ways meant to seem authoritative. And even when they do say truthful things, they bullshit — because they are *designed* to bullshit.[11]

Moving from a "hallucination" to a "bullshit" metaphor when talking about these AI models therefore has several practical benefits (Hicks et al. 2024, p. 37). It cautions us against overstating their abilities; it blunts worries about (or attempts to abdicate responsibility for) solving their inaccuracy problems; and it prompts appropriately sceptical attitudes even when they get things right. Speaking of AI gossip, we suggest, can have similar practical effects. It reminds us how fallible — and *feral* — these bullshit generators can be when it comes to evaluating and disseminating information. And it reinforces the fact that they are explicitly designed to elicit our trust and dependence, despite their unreliability and the various potential harms (both epistemic and technosocial) their misinformation can lead to.

We finish with a point signposted earlier. We indicated that our discussion of two kinds of AI gossip — *bot-to-user* and *bot-to-bot* — was not exhaustive. A third kind of gossip may soon be increasingly salient: *user-to-bot gossip*. In these cases, users might seed bots with different nuggets of gossip knowing the latter will, in turn, rapidly disseminate them in its characteristically feral way.[12] Bots might therefore act as intermediaries, responding to user-seeded gossip and rapidly spreading it to others (e.g., in response to inputs like "What do you think of Kevin Roose?") as well as to other bots (via integrated networks or shared databases and training data). This might be relatively benign, such as when a movie star's publicist seeds gossip about their client "anonymously" visiting a children's cancer ward shortly before the premiere of their latest film. However, there are many potentially dangerous use cases where bots might be weaponized by bad actors to spread gossip and inflict substantive technosocial harms on targeted individuals or groups.

---

[11] Researchers at Colombia's Tow Center for Digital Journalism recently asked ChatGPT to identify the source of two hundred quotes from twenty publications (including forty sources that disallowed OpenAI's search crawler to access their website). ChatGPT returned partially or entirely false information on one hundred and fifty-three occasions. More troublingly, it only acknowledged an inability to accurately respond to queries on seven occasions, when it used qualifying words like "appears", "it's possible", "might", or statements like "I couldn't locate the exact article" (Davis 2024).

[12] We're grateful to Samir Dayal, Arianna Falbo, and Adrian Currie for raising this point.

Once more, something like this already seems to be happening.[13] Recently, the Canadian Broadcasting Company found that suspected bots on social media and other media outlets favourable to Narendra Modi, the current Prime Minister of India, worked together to amplify misinformation about Canadian institutions and stoke tensions between Sikhs and Hindus in Canada (Montpetit et al. 2024). A handful of prominent Canadian influencers — known to be critical of the Khalistan movement (which advocates for an independent state for Sikhs and is opposed by pro-Modi communities promoting a Hindu nationalist ideology) — posted misleading and inflammatory content. This content included false claims about the significant role "pro-Khalistan extremists" supposedly play in Canadian Prime Minister Justin Trudeau's government, deceptively edited videos suggesting pro-Khalistan Sikh protestors were instigating violence (including attacks on Hindu temples), and other forms of rumour, innuendo, and gossip. This misinformation was quickly picked up and amplified by up to 1000 suspected AI bots, leading not only to heated discussions online but also to violent clashes outside Hindu temples and Sikh gurdwaras. It seems, then, that a small number of individuals successfully fed AI bots dangerous gossip with the intent to cause social instability and inflict real-world technosocial harms on specific groups. And it worked.

To conclude, AI offers both promise and peril. It can be an effective tool for learning and creating. But as a bullshit generator, it can be a source of misinformation and epistemic harm. We've argued that we should also be worried about something else: AI gossip and technosocial harm. As our examples show, this isn't a hypothetical threat. Real-world cases of AI gossip already exist, as does their resultant reputational damage, defamation, and social unrest. So, getting clear about these ideas isn't just a philosophical issue. These discussions have practical significance. Ideally, they will help us become more reflective, vigilant, and critical users, more sensitive to what this technology does and the scope of its potential benefit and harm.

---

[13] Many thanks to Tyler Brunet for drawing our attention to this example.

**References**

Adkins, K. (2017). *Gossip, epistemology, and power: Knowledge underground* (1st ed.) [PDF]. Springer International Publishing.

Alfano, M., & Robinson, B. (2017). Gossip as a burdened virtue. *Ethical Theory and Moral Practice: An International Forum*, *20*(3), 473–487.

Bansal, S., Garg, N., Singh, J., & Van Der Walt, F. (2023). Cyberbullying and mental health: past, present and future. *Frontiers in Psychology*, *14*, 1279234.

Bergstrom, C. T., & Ogbunu, C. B. (2023, April 6). *ChatGPT Isn't 'Hallucinating.' It's Bullshitting*. Undark Magazine. https://undark.org/2023/04/06/chatgpt-isnt-hallucinating-its-bullshitting/

Bhuiyan, J. (2024, December 14). She didn't get an apartment because of an AI-generated score – and sued to help others avoid the same fate. *The Guardian*. https://www.theguardian.com/technology/2024/dec/14/saferent-ai-tenant-screening-lawsuit

Birhane, A. (2021). Algorithmic injustice: a relational ethics approach. *Patterns (New York, N.Y.)*, *2*(2), 100205.

Birhane, A. (2022). The unseen Black faces of AI algorithms. *Nature*, *610*(7932), 451–452.

Bommasani, R., Hudson, D. A., Adeli, E., Altman, R., Arora, S., von Arx, S., Bernstein, M. S., Bohg, J., Bosselut, A., Brunskill, E., Brynjolfsson, E., Buch, S., Card, D., Castellon, R., Chatterji, N., Chen, A., Creel, K., Davis, J. Q., Demszky, D., … Liang, P. (2021). On the opportunities and risks of foundation models. In *arXiv [cs.LG]*. arXiv. http://arxiv.org/abs/2108.07258

Brooks, R. (2023). *I tried the Replika AI companion and can see why users are falling hard. The app raises serious ethical questions*. The Conversation. http://theconversation.com/i-tried-the-

replika-ai-companion-and-can-see-why-users-are-falling-hard-the-app-raises-serious-ethical-questions-200257

Buolamwini, J. (2023). *Unmasking AI: My mission to protect what is human in a world of machines*. Random House.

Davis, W. (2024, December 3). *ChatGPT's search results for news are 'unpredictable' and frequently inaccurate*. The Verge. https://www.theverge.com/2024/12/3/24312016/chatgpt-search-results-review-inaccurate-unpredictable

De Liban, K. (2024). *Inescapable AI: The Ways AI Decides How LowIncome People Work, Live, Learn, and Survive*. Techtonic Justice. https://www.techtonicjustice.org/reports/inescapable-ai

Dzieza, J. (2024, December 3). *What do you love when you fall for AI?* The Verge. https://www.theverge.com/c/24300623/ai-companions-replika-openai-chatgpt-assistant-romance

Edwards, B. (2023, April 6). *Why ChatGPT and Bing Chat are so good at making things up*. Ars Technica. https://arstechnica.com/information-technology/2023/04/why-ai-chatbots-are-the-ultimate-bs-machines-and-how-people-hope-to-fix-them/

Eickers, G. (2024). Social media experiences of LGBTQ+ people: Enabling feelings of belonging. *Topoi: An International Review of Philosophy*, *43*(3), 617–630.

Fisher, S. A. (2024). *Large language models and their big bullshit potential*. https://philpapers.org/rec/FISLLM

Fjelland, R. (2020). Why general artificial intelligence will not be realized. *Humanities & Social Sciences Communications*, *7*(1), 1–9.

Frankfurt, H. G. (2005). *On bullshit*. Princeton University Press.

Gabriels, K., & De Backer, C. J. S. (2016). Virtual gossip: How gossip regulates moral life in virtual worlds. *Computers in Human Behavior*, *63*, 683–693.

Goodman, R. F., & Ben-Ze'ev, A. (Eds.). (1994). *Good Gossip*. University Press of Kansas.

Haque, U. (2023). Architecture, Interaction, Systems. In B. Farahi & N. Leach (Eds.), *Interactive Design* (pp. 42–47). De Gruyter.

Harding, S. (2024, December 10). *Startup will brick $800 emotional support robot for kids without refunds*. Ars Technica. https://arstechnica.com/gadgets/2024/12/startup-will-brick-800-emotional-support-robot-for-kids-without-refunds/

Harper, G. W., Serrano, P. A., Bruce, D., & Bauermeister, J. A. (2016). The internet's multiple roles in facilitating the sexual orientation identity development of gay and bisexual male adolescents. *American Journal of Men's Health*, *10*(5), 359–376.

Hartung, F.-M., Krohn, C., & Pirschtat, M. (2019). Better than its reputation? Gossip and the reasons why we and individuals with "dark" personalities talk about others. *Frontiers in Psychology*, *10*, 1162.

Heersmink, R., de Rooij, B., Vázquez, M. J. C., & Colombo, M. (2024). A phenomenology and epistemology of large language models: transparency, trust, and trustworthiness. *Ethics and Information Technology*, *26*, 41.

Jolly, E., & Chang, L. J. (2021). Gossip drives vicarious learning and facilitates social connection. *Current Biology: CB*, *31*(12), 2539-2549.e6.

Kaye, B. (2023, April 5). Australian mayor readies world's first defamation lawsuit over ChatGPT content. *Reuters*. https://www.reuters.com/technology/australian-mayor-readies-worlds-first-defamation-lawsuit-over-chatgpt-content-2023-04-05/

Krueger, J., & Osler, L. (2019). Engineering Affect: Emotion Regulation, the Internet, and the Techno-Social Niche. *Philosophical Topics*, *47*(2), 205–231.

Krueger, J., & Osler, L. (2022). Communing with the Dead Online: Chatbots, Grief, and Continuing Bonds. *Journal of Consciousness Studies*, *29*(9–10), 222–252.

Krueger, J., & Roberts, T. (2024). Real feeling and fictional time in human-AI interactions. *Topoi: An International Review of Philosophy*, *43*(3), 783–794.

Lind, P. G., da Silva, L. R., Andrade, J. S., Jr, & Herrmann, H. J. (2007). Spreading gossip in social networks. *Physical Review. E, Statistical, Nonlinear, and Soft Matter Physics*, *76*(3 Pt 2), 036117.

Maples, B., Cerit, M., Vishwanath, A., & Pea, R. (2024). Loneliness and suicide mitigation for students using GPT3-enabled chatbots. *Npj Mental Health Research*, *3*(1), 1–6.

McMahon, L., & Kleinman, Z. (2024, May 24). *Google AI search tells users to glue pizza and eat rocks*. BBC News. https://www.bbc.co.uk/news/articles/cd11gzejgz4o

Merry, S. E. (1984). Rethinking Gossip and Scandal. In *Toward a General Theory of Social Control* (pp. 271–302). Elsevier.

Montpetit, J., Bhugra, S., & Angelovski, I. (2024, December 18). Bots and Indian TV push fake news about Canada in wake of Hindu temple clashes. *CBC News*. https://www.cbc.ca/news/canada/toronto/bots-temple-clashes-hindu-sikh-canada-1.7411094

Obermeyer, Z., Powers, B., Vogeli, C., & Mullainathan, S. (2019). Dissecting racial bias in an algorithm used to manage the health of populations. *Science (New York, N.Y.)*, *366*(6464), 447–453.

OK_Surprise_7973}. (2024, December 18). *ChatGPT is one of my best friends right now and I'm tired of pretending it's not*. Reddit. https://www.reddit.com/r/OpenAI/comments/1hgurbx/chatgpt_is_one_of_my_best_friends_right_now_and/?share_id=WeIFdrvUJKQA-AmEanKBc&utm_content=2&utm_medium=ios_app&utm_name=iossmf&utm_source=share&utm_term=22
26

Orland, K. (2024, September 25). *Talking to ChatGPT for the first time is a surreal experience*. Ars Technica. https://arstechnica.com/ai/2024/09/talking-to-chatgpt-for-the-first-time-is-a-surreal-experience/

Osler, L., & Krueger, J. (2022). Taking Watsuji online: betweenness and expression in online spaces. *Continental Philosophy Review*, *55*(1), 77–99.

Pika, J. (2019, July 10). *Queer Youth Exploring Their Identity, One Webpage at a Time*. Center for the Study of Social Policy. https://cssp.org/2019/07/queer-youth-exploring-identity-online/

Piltch, A. (2024, May 25). *17 cringe-worthy Google AI answers demonstrate the problem with training on the entire web*. Tom's Hardware. https://www.tomshardware.com/tech-industry/artificial-intelligence/cringe-worth-google-ai-overviews

Radzik, L. (2016). Gossip and social punishment. *Res Philosophica*, *93*(1), 185–204.

Ratan, Z. A., Parrish, A.-M., Zaman, S. B., Alotaibi, M. S., & Hosseinzadeh, H. (2021). Smartphone addiction and associated health outcomes in adult populations: A systematic review. *International Journal of Environmental Research and Public Health*, *18*(22), 12257.

Roose, K. (2023, February 16). A Conversation With Bing's Chatbot Left Me Deeply Unsettled. *The New York Times*. https://www.nytimes.com/2023/02/16/technology/bing-chatbot-microsoft-chatgpt.html

Roose, K. (2024a, August 30). How Do You Change a Chatbot's Mind? *The New York Times*. https://www.nytimes.com/2024/08/30/technology/ai-chatbot-chatgpt-manipulation.html

Roose, K. (2024b, December 13). How Claude Became Tech Insiders' Chatbot of Choice. *The New York Times*. https://www.nytimes.com/2024/12/13/technology/claude-ai-anthropic.html

Shevlin, H. (2024). All too human? Identifying and mitigating ethical risks of Social AI. *Law, Ethics & Technology*, *1*(2), 1–22.

Slater, J., Humphries, J., & Hicks, M. T. (2024, July 17). *ChatGPT Isn't 'Hallucinating'—It's Bullshitting!* Scientific American. https://www.scientificamerican.com/article/chatgpt-isnt-hallucinating-its-bullshitting/

Solove, D. J. (2007). *The future of reputation: Gossip, rumor, and privacy on the internet*. Yale University Press.

Turley, J. (2023, April 3). ChatGPT falsely accused me of sexually harassing my students. Can we really trust AI? *USA Today*. https://www.usatoday.com/story/opinion/columnist/2023/04/03/chatgpt-misinformation-bias-flaws-ai-chatbot/11571830002/

van Niekerk, J. (2008). The virtue of gossip. *South African Journal of Philosophy*, *27*(4), 400–412.

Vincent, J. (2023, June 9). *OpenAI sued for defamation after ChatGPT fabricates legal accusations against radio host*. The Verge. https://www.theverge.com/2023/6/9/23755057/openai-chatgpt-false-information-defamation-lawsuit

Wertheimer, T. (2022, July 23). *Blake Lemoine: Google fires engineer who said AI tech has feelings*. BBC News. https://www.bbc.co.uk/news/technology-62275326